\documentclass[epj]{svjour}
\usepackage{graphics,epsfig}

\begin{document}
\title{Longitudinal broadening of near side jets due to parton cascade}
\titlerunning{Longitudinal broadening of near side jets due to parton cascade}
\author{
G. L. Ma\inst{1} \and S. Zhang \inst{1,2} \and Y. G.
Ma\inst{1}\thanks{Corresponding author: Email: ygma@sinap.ac.cn}
 \and
X. Z. Cai \inst{1} \and J. H. Chen \inst{1,2} \and C.
Zhong\inst{1} }

\institute{ Shanghai Institute of Applied Physics, Chinese Academy
of Sciences, P.O. Box 800-204, Shanghai 201800, China \and
Graduate School of the Chinese Academy of Sciences, Beijing
100080, China }
%
\date{Received: date / Revised version: date}
%
\abstract{
Longitudinal broadening along $\Delta\eta$ direction on near side in two-dimensional ($\Delta\phi \times \Delta\eta$) di-hadron correlation distribution has been studied for central Au+Au collisions at $\sqrt{s_{NN}}$ = 200 GeV, within a dynamical multi-phase transport model. It was found that the longitudinal broadening is generated by a longitudinal flow induced by strong parton cascade in central Au+Au collisions, in comparison with p+p collisions at $\sqrt{s_{NN}}$ = 200 GeV. The longitudinal broadening may shed light on  the information about strongly interacting partonic matter at RHIC.
\PACS{ {25.75.Nq}{Quark deconfinement} \and {25.75.Gz}{Particle correlations and fluctuations}
     } 
}

\maketitle

\section{Introduction}
   \label{intro}
Di-hadron correlation is thought as a good probe to explore the
properties of  strongly interacting partonic matter in central
Au+Au collisions at the Relativistic Heavy Ion Collider
(RHIC)~\cite{White-papers}. In di-hadron azimuthal correlations,
an away side jet is found quenching and losing its energy into the
nearby dense medium, since it has a long interactive path in the
medium, i.e. jet quenches~\cite{HIJING}, which has been supported
by many experimental observations such as the medium  modification
of away side jets in di-hadron azimuthal
correlation~\cite{hard-hard-ex,sideward-peak2}. On the other hand,
would a near side jet emitted near the surface be affected by the
formation of dense partonic matter?  The recently observed
longitudinal broadening sitting at a plateau on near side in the
$\Delta\eta$ direction of two-dimensional ($\Delta\phi \times
\Delta\eta$) di-hadron correlation function, so-called $'ridge'$
correlation~\cite{ridge,fuqiangPRL,ridgeqm06}, is attracting the
attentions from theorists (where $\Delta\phi$ and $\Delta\eta$
represent the pseudo-rapidity and azimuthal angle differences
between the trigger and associated particles), respectively. Chiu
and Hwa concluded that it results from enhanced thermal partons by
the energy loss of hard partons traversing the bulk medium by
using a recombination model~\cite{chiu}. Armesto et
al.~\cite{Armesto} and Satarov et al.~\cite{Sat} explained it as a
consequence of the longitudinal flow of medium created at early
stages of a heavy-ion collision. Shuryak argued that the ridge is
produced by QCD bremsstrahlung along the beam and then boosted by
transverse flow~\cite{ShuryakRidge}. Majumder and M\"uller et al.
proposed that turbulent color fields caused by plasma
instabilities can deflect the transversely propagating partons in
the direction of beam axis~\cite{BMuller-ridge}. Romatschke
presented a longitudinal broadening in the frame of elastic
collisional energy loss of heavy quarks in an anisotropic
expanding system~\cite{Romatschke}. Wong simulated the ridge
structure by a momentum kick model in which ridge particles are
identified as medium partons which suffer collisions with the jet
and acquire momentum kicks along the jet direction~\cite{Wong}.
However, Pantuev thought it could be the result of the back splash
from stopped energetic partons~\cite{Pantuev}.  In this paper, by
using a dynamical multi-phase transport model, we found that the
longitudinal flow in partonic longitudinal transport process can
generate the longitudinal broadening with the evolution of
partonic matter in central Au+Au collisions at $\sqrt{s_{NN}}$=200
GeV, in comparison with the results of p+p collisions at
$\sqrt{s_{NN}}$ = 200 GeV.

\section{Model Introduction}
  \label{Model}

A multi-phase transport model (AMPT) model \cite{AMPT}  is a
hybrid model which consists of four main processes: the initial
conditions, partonic interactions, the conversion from partonic
matter into hadronic matter and hadronic interactions. The initial
conditions, which include the spatial and momentum distributions
of minijet partons and soft string excitations, are obtained from
the HIJING model~\cite{HIJING}. The parton structure functions are
taken as Duke-Owen structure function set 1~\cite{Duk82}. A $K$
factor of 2.5 has been included to correct the contribution from
lowest order pQCD processes. Excitations of strings  melt strings
into partons. Scatterings among partons are modelled by Zhang's
parton cascade model  (ZPC) \cite{ZPC}, which at present includes
only two-body scattering with cross section obtained from the pQCD
with screening mass. In the default version of the AMPT model
\cite{DAMPT}(abbr. ``the default AMPT" model) partons are
recombined with their parent strings when they stop interactions,
and the resulting strings are converted to hadrons by using a Lund
string fragmentation model \cite{Lund}. In  the string melting
version of the AMPT model (abbr. ``the melting AMPT"
model)\cite{SAMPT}, a simple quark coalescence model is used to
combine partons into hadrons. Dynamics of the subsequent hadronic
matter is then described by A Relativistic Transport (ART) model
\cite{ART}. Details of the AMPT model can be found in a recent
review \cite{AMPT}. As shown in previous studies about elliptic
flow~\cite{AMPT,SAMPT,Jinhui} and Mach cone-like structure
\cite{glmaPLB}, the partonic effect can not be neglected in the
relativistic heavy-ion collisions. Therefore, the melting AMPT
model is much more appropriate than the default AMPT model,
especially when the energy density is much higher than the
critical density for the QCD phase transition. But in these
previous studies with the melting AMPT
model~\cite{AMPT,SAMPT,Jinhui,glmaPLB}, parton cascade process
 does not stop until
partons cease to interact, and the hadronization takes place
dynamically during the process. It is not reasonable since
partonic matter has a limited evolution-time which depends on when
the energy density (temperature) of reaction system enters the
critical threshold, in a Lattice QCD context~\cite{LatticeQCD}. In
this study, we tested different evolution-times of partonic
matter, after which no partonic interactions are allowed and all
left over partons must coalesce into hadrons suddenly. It means
that the partonic matter comes to the critical point at a certain
time, therefore the evolution-time of partonic matter is
comparable with the period during which partons interact. We
simulated central Au+Au collisions (0-10\% centrality) at
$\sqrt{s_{NN}}$=200 GeV with the melting AMPT model and p+p
collisions at $\sqrt{s_{NN}}$=200 GeV with the default AMPT model.
The partonic interacting cross section in the melting AMPT model
is  taken as 10 mb.

\section{Analysis Method}
  \label{anamethod}

Two-dimensional ($\Delta\phi \times \Delta\eta$) di-hadron
correlation functions between the trigger and associated hadrons
were constructed by a mixing-event technique in our analysis. Here
$\Delta\phi = \phi - \phi_{trig}$ and $\Delta\eta = \eta -
\eta_{trig}$ represents the difference of azimuthal angle and
pseudo-rapidity between the trigger and associated particles,
resepctively. The  $p_{T}$ window cuts for trigger and associated
particles are $2.5 < p_{T}^{trig} < 6$ GeV/$c$ and $1.5 <
p_{T}^{assoc} < 2.5$ GeV/$c$, respectively. Both trigger and
associated particles were selected within pseudo-rapidity window
$|\eta| < 1.0$. The pairs of the associated particles with trigger
particles in the same events were accumulated to obtain a
$\Delta\phi \times \Delta\eta$ distribution. In order to remove
the background which is expected to mainly come from the effect of
elliptic flow \cite{soft-soft-ex,sideward-peak2}, so-called
mixing-event method was applied to simulate the background. In
this method, we mixed two events which have very close impact
parameters (the impact parameter difference of two events is
required to be less than 0.5 fm) and the same reaction plane
direction into a mixed event, and extracted $\Delta\phi \times
\Delta\eta$ distribution which is regarded as the corresponding
background. A zero yield at minimum (ZYAM)
assumption~\cite{sideward-peak2} was adopted to subtract the
background by making the associated yield be zero around
$|\Delta\phi|$=1 (i.e. the strip of $0.9< |\Delta\phi| < 1.1$)
after substraction.

\section{Results}
 \label{results}

\begin{figure}
\includegraphics[scale=0.45]{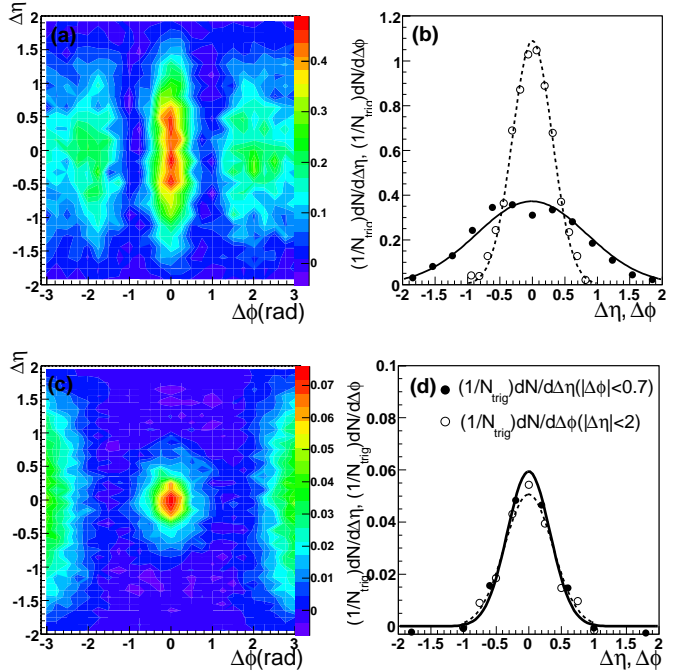}
\caption{\footnotesize (Color online) Two-dimensional
 ($\Delta\phi \times \Delta\eta$) di-hadron
 correlation functions (a) and the projected $\Delta\eta$
 (full circles, within $|\Delta\phi|$ $<$ 0.7) and
 $\Delta\phi$ (open circles, within $|\Delta\eta|$ $<$ 2)
  distributions of associated hadrons (b) on near side for
  central Au+Au collisions (0-10\%) in the melting AMPT model
   with hadronic rescattering; (c) and (d): The corresponding
   ones for p+p collisions at $\sqrt{s_{NN}}$=200 GeV in the
   default AMPT model with hadronic rescattering. The lines
   in (b) and (d) are the Gaussian fitting
   functions to the $\Delta\eta$ (solid) and $\Delta\phi$ (dash) projections.
 }
\label{AuAudAu}
\end{figure}

Figure~\ref{AuAudAu}(a) and (c) give the background subtracted two-dimensional ($\Delta\phi \times \Delta\eta$)
di-hadron correlation functions for central Au+Au collisions (0-10\%) at $\sqrt{s_{NN}}$=200 GeV in the melting
AMPT model with hadronic rescattering and p+p collisions at $\sqrt{s_{NN}}$=200 GeV in the default AMPT model
with hadronic rescattering, respectively. A significant longitudinal broadening in the $\Delta\eta$ direction on
near side ($|\Delta\phi| < 1.0$) is observed in central Au+Au collisions, as compared to that in p+p collisions.
Figure 1(b) and 1(d) gives the projected $\Delta\eta$
 (within $|\Delta\phi| < 0.7$) and $\Delta\phi$ (within $|\Delta\eta|$ $<$ 2)
  distributions of associated hadrons for near side in Au+Au and p+p collisions
  respectively, where the solid and dash lines give the corresponding Gaussian
  fitting functions.  As a result, the ratio between two root mean squares of the
  projected $\Delta\eta$ and  $\Delta\phi$ distributions (
i.e. rms($\Delta\eta$)/ rms($\Delta\phi$)) was found to be
$\sim$2.6 in central Au+Au collisions, but $\sim$1.0 in p+p
collisions, which is consistent with the preliminary experimental observation~\cite{ridge} and theoretical calculation~\cite{Romatschke}.

\begin{figure}
\includegraphics[scale=0.45]{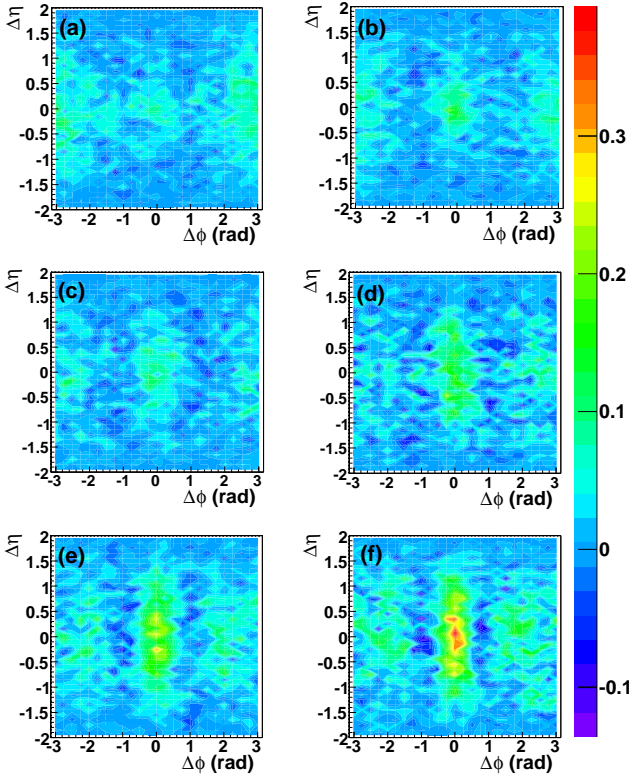}
\caption{\footnotesize (Color online) Two-dimensional ($\Delta\phi \times \Delta\eta$)di-hadron
correlation functions with different evolution-times of partonic matter in central Au+Au collisions (0-10\%) at
$\sqrt{s_{NN}}$=200 GeV in melting AMPT model without hadronic rescattering. (a): 0.5 fm/c; (b): 1.0 fm/c; (c):
1.5 fm/c; (d): 2.5 fm/c; (e): 3.0 fm/c; (f): 4.0 fm/c.
 }
\label{evolution}
\end{figure}

In order to learn about the evolution of longitudinal broadening
in the parton cascade processes for central Au+Au collisions,
Fig.~\ref{evolution}(a)-(f) presents the two-dimensional
($\Delta\phi \times \Delta\eta$) di-hadron correlation functions
with different evolution-times of partonic matter, where the
effect of hadronic rescattering is found to be negligible in the
study.
 As seen from figure~\ref{evolution} (a) and (b),
no obvious broadening is seen in the longitudinal $\Delta\eta$
direction with a short evolution-time. But the longitudinal broadening
emerges and grows up with the evolution-time ((c)-(f)), which
indicates strong and continuous parton cascade processes broaden
the $\Delta\eta$ distribution of the associated
particles on near side longitudinally. On the other hand,  near side seems
unchanged with partonic evolution-time in shape in the
$\Delta\phi$ direction.

\begin{figure}
\includegraphics[scale=0.40]{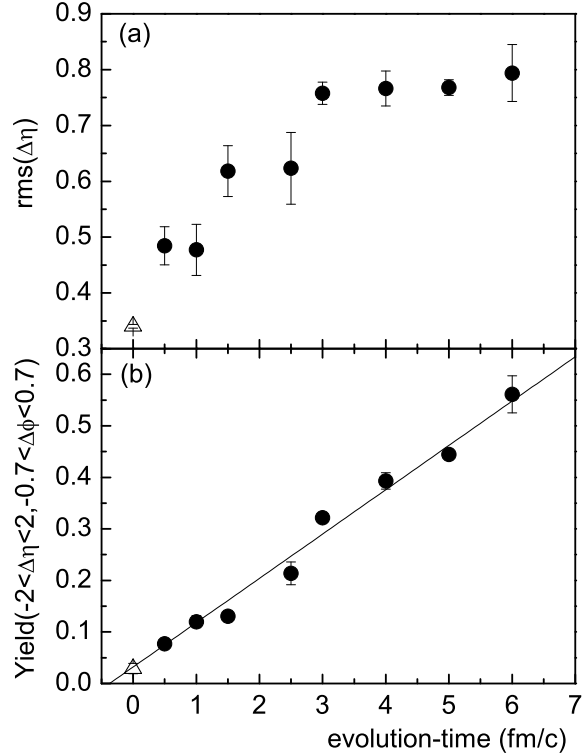}
\caption{\footnotesize  Full circles: the evolution-time dependences of
rms($\Delta\eta$) (a) and integrated near side yield  within
$|\Delta\phi|$ $<$ 0.7 (b) in Au+Au collisions at
$\sqrt{s_{NN}}$=200 GeV in the melting AMPT model without hadronic
rescattering; Open triangles: the corresponding values in p+p
collisions at $\sqrt{s_{NN}}$=200 GeV without hadronic
rescattering; Solid line in (b) shows a linear fitting function,
see the text for the detail.
 }
\label{rmsyield}
\end{figure}

Quantitatively, figure~\ref{rmsyield} (a) and (b) give the
evolution-time dependences  of rms ($\Delta\eta$) and the
integrated yield for near side within $|\Delta\phi| < 0.7$ in
central Au+Au collisions (0-10\%) at $\sqrt{s_{NN}}$ = 200 GeV,
respectively, where the final hadronic rescattering has not been
included. Because p+p collisions are thought to rarely experience
the processes of parton cascade commonly, the results of  p+p
collisions at $\sqrt{s_{NN}}$ = 200 GeV are plotted at zero
evolution-time  for the reference. From figure~\ref{rmsyield}(a),
it is apparent that the rms($\Delta\eta$) of near side increases
and saturates with the evolution-time of partonic matter. It can
also be seen in figure~\ref{rmsyield}(b), the associated near-side
yield per trigger particle seems to increase linearly with the
evolution-time (the slope= 0.085 $\pm$ 0.002 $fm^{-1}$), which is
consistent with our previous work~\cite{evolutionpaper}.

\section{Discussions}
 \label{diss}

\begin{figure}
\includegraphics[scale=0.40]{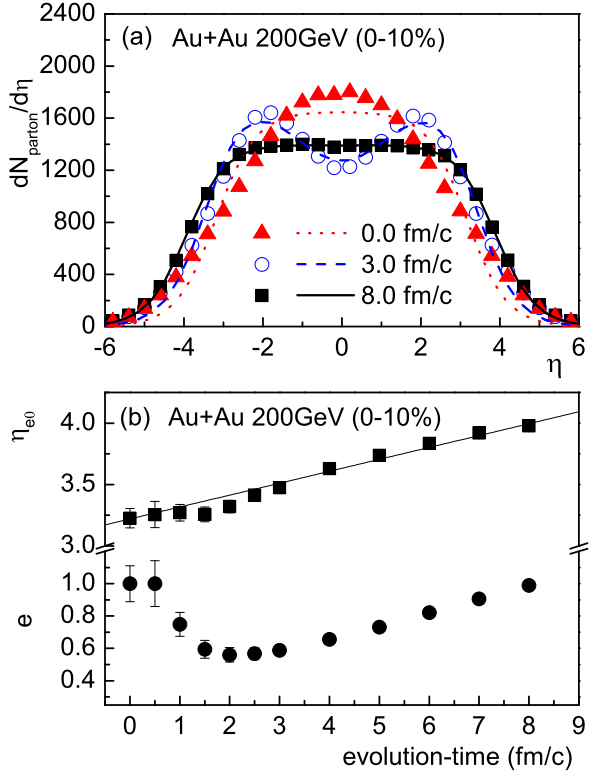}
\caption{\footnotesize  (Color online) (a) The pseudo-rapidity
($\eta$) distributions of partons with three different
evolution-times of partonic matter in  Au+Au collisions at
$\sqrt{s_{NN}}$=200 GeV in the melting AMPT model (points), where
the lines shows the fits from NUFM model which is defined in the
text and Ref.~\cite{NUFM}; (b) The ellipticity e (circles) and the
rapidity limit $\eta_{e0}$ (squares), from NUFM fits to the $\eta$
distributions of partons, as a function of the evolution-time of
partonic matter in Au+Au collisions at $\sqrt{s_{NN}}$=200 GeV in
the melting AMPT model, where solid line shows a linear fitting
function.
 }
\label{raps}
\end{figure}

To investigate partonic longitudinal transport process in central
Au+Au collisions, figure~\ref{raps} (a) presents the
pseudo-rapidity distributions of partons at different partonic
evolution-times in central Au+Au collisions at $\sqrt{s_{NN}}$ =
200 GeV. We used a non-uniform longitudinal flow model
(NUFM)~\cite{NUFM} to fit these partonic pseudo-rapidity
distributions from AMPT model, as shown by the lines in
figure~\ref{raps} (a).   As a consequence, two fitting parameters
(i.e.  ellipticity $e$ (circles) and rapidity limit $\eta_{e0}$
(squares)) as a function of the evolution-time of  partonic matter
are gotten in figure~\ref{raps} (b). The ellipticity, which
represents the uniformity of fireball in the longitudinal
direction, drops until 3.0 fm/c firstly and then increases with
the evolution-time, which implies the scenario that the
longitudinal evolution starts from the superposition of projectile
and target systems, and they penetrate each other and diffuse via
further parton interactions. On the other hand,  $\eta_{e0}$,
which represents the amplitude of longitudinal flow, increases
with evolution-time almost linearly (the slope = 0.093 $\pm$ 0.003
$fm^{-1}$, which is consistent with that of  the associated
near-side yield shown in figure~\ref{rmsyield}(b)).  Therefore,
the existence of strong partonic longitudinal flow can result in
the above observed elongated structure in the $\Delta\eta$
direction.

In Ref~\cite{chiu}, a hard parton near surface was predicted to
lose some amount of energy to enhance thermal motion of the
partons in the environment, which was expected to result in the
longitudinal broadening of near side in central Au+Au collisions
at RHIC energy. However it has not given the dynamical detail
about how a hard parton losses its energy into the medium. In
Ref~\cite{Romatschke}, Romatschke discussed the collisional energy
loss of hard parton is in favor of the production of the
longitudinal broadening, but in the frame of heavy quarks. In the
present work, we found that the longitudinal broadening can be
produced by the longitudinal flow induced by parton cascade
longitudinally, by using a dynamical multi-phase transport model.
On the other hand, we presented the width of longitudinal
broadening (rms($\Delta\eta$)) increases with the evolution-time
of partonic matter , which is consistent with the conclusion in
Ref~\cite{Sat} that the width of elongated structure is
proportional to the logarithmic $'formation$ $time'$ of near side
correlation for a longitudinally-expanding QGP with a constant
sound velocity.

Recently, three-particle $\Delta\eta$ correlation is expected to
play a role to identify the possible physical mechanisms on
$'ridge'$~\cite{pawan}, and our corresponding work is in progress.
Though the plateau of $'ridge'$ correlation is not  observed in
our model, our present work has provided one possibility to
interpret  longitudinal broadening (i.e. induced by strong parton
cascade) on near side which was observed in central Au+Au
collisions at RHIC energy.

\section{Summary}
  \label{summ}

To summarize, we have explored the origin of longitudinal
broadening on near side in central Au+Au collisions at RHIC energy
within the framework of multi-phase transport model. In comparison
with p+p collisions, it has been found that the longitudinal
broadening in central Au+Au collisions is driven by the
longitudinal flow induced by strong parton cascade. Therefore the
researches on longitudinal broadening can give more information
about early partonic stage in relativistic heavy ion collisions.

\begin{acknowledgement}
Authors thank  Prof H. Z. Huang for the discussion. This work was
supported in part by the National Natural Science Foundation of
China  under Grant No 10610285 and 10705044, and  the Knowledge
Innovation Project of Chinese Academy of Sciences under Grant No.
KJCX2-YW-A14 and KJXC3-SYW-N2, the Startup Foundation for the CAS
Presidential Scholarship Award under Grant No. 29010702.
\end{acknowledgement}

\end{document}